\documentclass[epjST]{svjour}
\usepackage[sort&compress,numbers]{natbib}
\usepackage{amsfonts}
\usepackage{amsmath}
\usepackage{amssymb}
\usepackage{array}
\usepackage{bbold}
\usepackage{bm}
\usepackage{booktabs}
\usepackage{dsfont}
\usepackage[dvipsnames,usenames]{color}
\usepackage{enumitem}
\usepackage{float}
\usepackage{graphicx}
\usepackage{lipsum}
\usepackage{multirow}
\usepackage{nicefrac}
\usepackage[pdftex,
            pdftitle={3b-review},
            pdfauthor={Maxim Mai, Michael Doring, Akaki Rusetsky},
            bookmarks,
            colorlinks,
            linkcolor=blue,
            citecolor=blue,
            menucolor=black,
            urlcolor=blue,
            plainpages=false,
            pdfpagelabels,
            hypertexnames=false]{hyperref}
\usepackage[normalem]{ulem}
\usepackage[dvipsnames]{xcolor}

\setlist[enumerate]{leftmargin=30pt, itemsep=0pt}
\setlist[description]{leftmargin=40pt}

\numberwithin{equation}{section}

\setlist[itemize]{leftmargin=10pt, itemsep=5pt, label={\bf--}}
\graphicspath{{./pics/}}

\begin{document}

\title{Multi-particle systems on the lattice and chiral extrapolations: a brief review}

\author{Maxim Mai\inst{1}\fnmsep\thanks{\email{maximmai@gwu.edu}}
\and Michael D\"oring\inst{1,2}\fnmsep\thanks{\email{doring@gwu.edu}}
\and Akaki Rusetsky\inst{3,4}\fnmsep\thanks{\email{rusetsky@hiskp.uni-bonn.de}}
}

\institute{The George Washington University, Washington, DC 20052, USA 
\and Thomas Jefferson Accelerator Facility, Newport News, VA 23606, USA
\and HISKP and BCTP, Rheinische Friedrich-Wilhelms-Universit\"at Bonn,\\ 53115 Bonn, Germany
\and Tbilisi State University, 0186 Tbilisi, Georgia
}

\abstract{The extraction of two- and three-body hadronic scattering
amplitudes and the properties of the low-lying hadronic resonances from the
finite-volume energy levels in lattice QCD represents a rapidly developing field of
research. The use of various modifications of the L\"uscher finite-volume method has
opened a path to calculate infinite-volume scattering amplitudes on the lattice. Many
new results have been obtained recently for different two- and three-body scattering
processes, including the extraction of resonance poles and their properties from lattice
data. Such studies, however, require robust parametrizations of the infinite-volume
scattering amplitudes, which rely on basic properties of $S$-matrix theory and --
preferably -- encompass systems with quark masses at and away from the physical point. Parametrizations of this kind, provided by unitarized Chiral Perturbation Theory, are discussed in this  review. Special attention is paid to three-body systems on the lattice, owing to the rapidly growing interest in the field. Here, we briefly survey the formalism, chiral extrapolation, as well as finite-volume analyses of lattice data.
{\bf Report number:} JLAB-THY-21-3327
}

\maketitle   

\clearpage

\section{Introduction}
\label{sec:intro-mm}

Strong interactions govern the formation of protons, neutrons, and nuclei.
Scattering and decay experiments provide access to strong interaction phenomena,
implying that theoretical approaches should describe amplitudes which involve two or more
asymptotically stable states. Quantum Chromodynamics (QCD) on the lattice represents a
framework for the {\it ab initio} access to such multiparticle amplitudes. However,
pertinent calculations are performed at Euclidean times that allows a direct extraction
of scattering amplitudes only at threshold~\cite{Maiani:1990ca}. Furthermore, calculations
of lattice QCD (LQCD) are performed in small boxes. Since the spectrum in a finite volume is discrete,
it is clear that, on the lattice, one does not have direct access to the scattering amplitudes. However, in his groundbreaking
papers~\cite{Luscher:1985dn, Luscher:1986pf, Luscher:1990ux}, L\"uscher has shown
that the quantization of the energy levels in a finite box can be turned into an advantage
that allows one to circumvent the no-go theorem of Ref.~\cite{Maiani:1990ca}. In particular, the
two-body energy levels, measured on the Euclidean lattice, can be directly mapped onto
the two-body elastic scattering phase shift, which is defined in the infinite-volume
Minkowski space. This novel idea paved the way to  studies of scattering
processes in lattice QCD, which have gained much popularity. The approach has
been subject to different generalizations, including the application to study resonance
decays and form factors, as well as the extension to three- and more particles. In the
present review, we address some of these developments.

\begin{figure}
    \centering
    \includegraphics[width=0.40\linewidth]{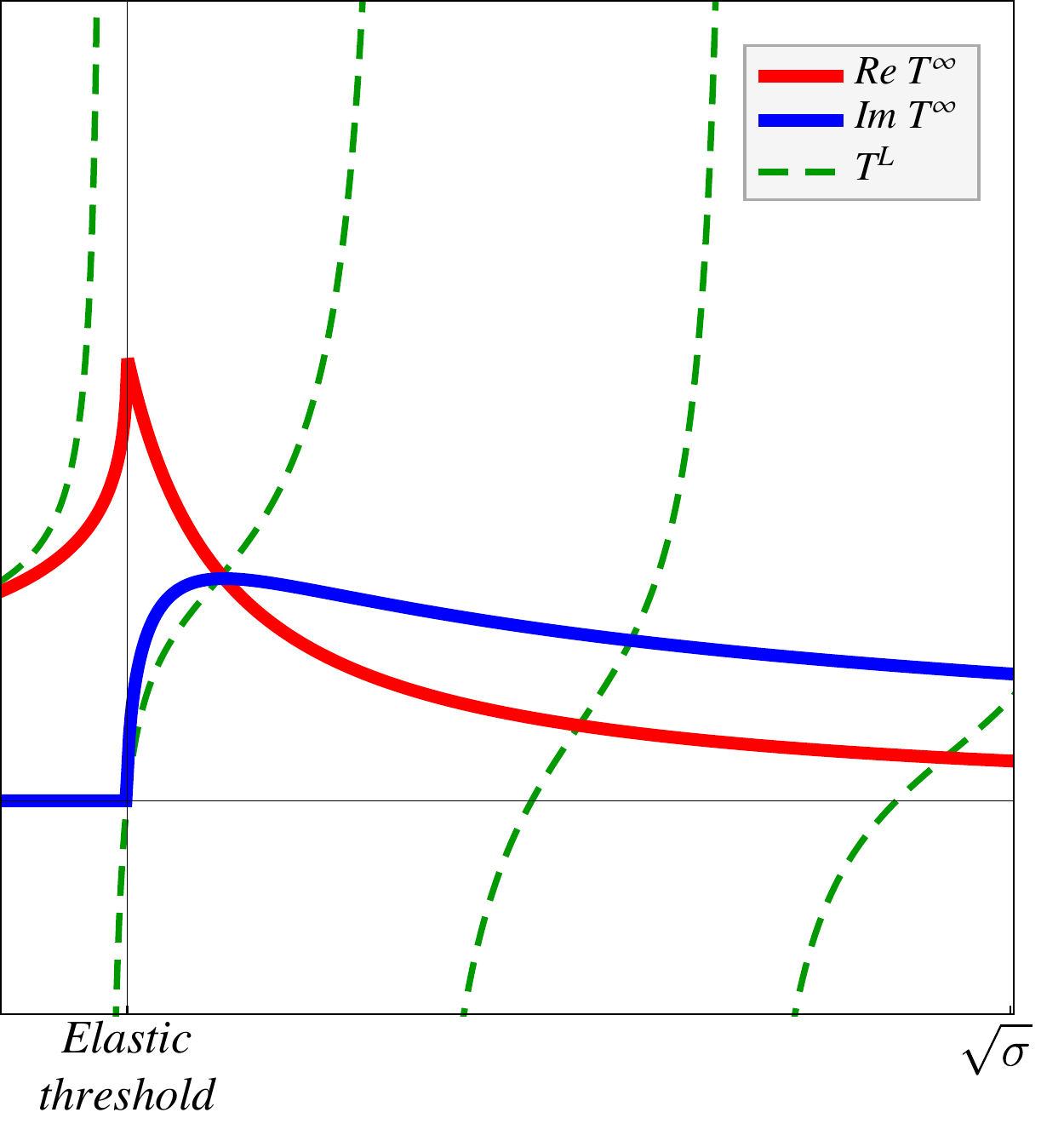}
    \hspace*{0.4cm}
    \includegraphics[width=0.54\linewidth,trim= 0 3cm 12cm 7.2cm,clip]{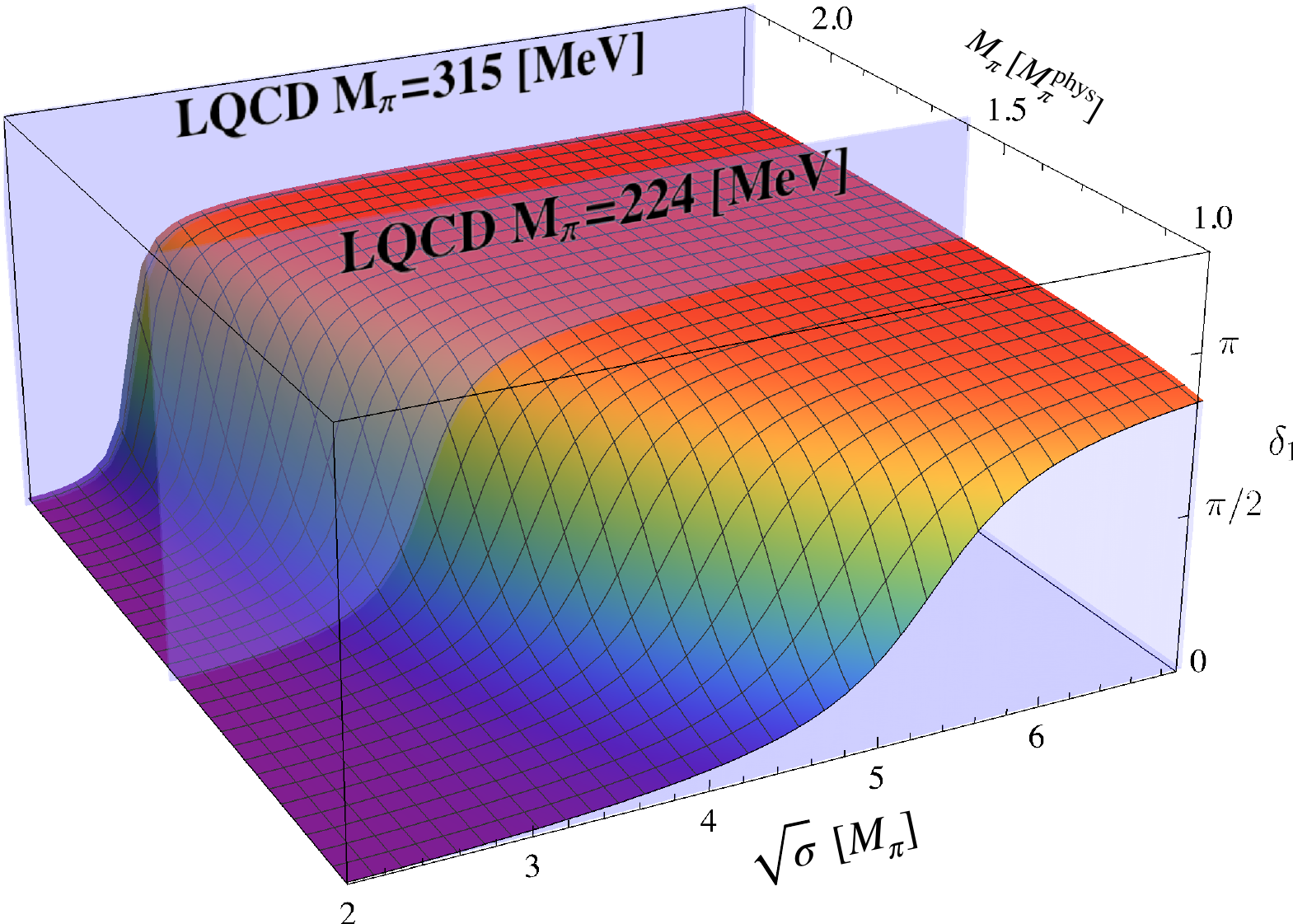}
    \caption{
{\bf Left:}
A schematic representation of the S-wave scattering amplitude in a scalar field theory as a function of two-body energy $\sqrt{\sigma}$, calculated in the infinite and and in a finite volume. Whereas the amplitude $T^\infty$ is smooth apart from threshold and complex, the real-valued finite-volume amplitude $T^L$ has simple poles at the energies corresponding to the spectrum of the Hamiltonian in a finite volume; {\bf Right:} An example of the interplay between lattice QCD and ChPT in describing the $\pi\pi$ phase shift in the vicinity of the $\rho$-resonance. ChPT connects phase shifts at different energies and pion masses. The fixed-$M_\pi$ planes contain constraints from experiment and lattice QCD calculations~\cite{Guo:2016zos, Mai:2019pqr}. 
}
    \label{fig:fin-inf--chiral-concepts}
\end{figure}

Hadronic systems are accessed in LQCD by calculating correlation functions on a
discretized Euclidean space-time in a finite volume. Thus, the ``raw'' lattice results should
be corrected for different lattice artifacts before a comparison to the real world can be
made. Effective field theory methods can be used to treat each of these artifacts. First,
there are the so-called discretization effects that are linked to the finite lattice spacing $a$.
A continuum limit $a\to0$ needs to be performed to relate lattice QCD results to 
physical quantities. In addition, one needs to establish a connection to physical units,
referred to as scale-setting. Both these issues are related to practical aspects of
LQCD calculations and are beyond the scope of the present review. More importantly, LQCD calculations are necessarily carried out in a finite volume.
Consider for simplicity a spatially\footnote{The Euclidean time dimension does not
  play any role in this review and will be always assumed to be infinite.} cubic lattice with side length $L$. Imposing boundary conditions in the spatial directions leads to the
quantization of the three-momenta of particles. For example, in case of periodic boundary
conditions, the allowed momenta are ${\bm p}=2\pi{\bm n}/L$ for
${\bm n}\in \mathds{Z}^3$. The spectrum in a finite volume is discrete as well, and
the position of the energy levels depends on $L$. Then, it can
be verified that the matrix elements, given by the sum of all Feynman diagrams, calculated
in a finite volume, exhibit an irregular behavior in $L$. This is illustrated in the left panel of
Fig.~\ref{fig:fin-inf--chiral-concepts} which schematically shows a
two-particle S-wave elastic amplitude, calculated in infinite and finite volume
(in the latter case, the integration over three-momenta in the Feynman integrals is
 replaced by the summation over the discrete values). As seen from this figure,
the amplitude in the infinite volume is a complex-valued smooth function of the energy.
On the contrary, the amplitude in a finite volume is real and discontinuous: it has
first-order poles at the energies corresponding to the discrete energy eigenvalues
for a given $L$. Clearly, increasing $L$ leads to the condensation of singularities on the
real axis, but the limit $L\to\infty$ is not well defined.
Hence, in a finite volume, one first needs to identify the quantities which exhibit smooth
behavior for large $L$, and perform the limit $L\to \infty$ for these quantities only.
This is  the essence of L\"uscher's method~\cite{Luscher:1984xn}.

Over time, extensions of the two-body L\"uscher formalism to moving frames and
modified box geometries have been formulated~\cite{Rummukainen:1995vs, Li:2003jn, Feng:2004ua, Kim:2005gf, Lage:2009zv, Fu:2011xz, Davoudi:2011md, Doring:2012eu, Leskovec:2012gb,  Briceno:2012yi, Gockeler:2012yj, Guo:2012hv, Li:2012bi, Briceno:2013hya, Lee:2017igf, Morningstar:2017spu, Li:2019qvh}. These techniques mostly serve to calculate more energy eigenvalues in the elastic scattering energy interval to better sample the amplitude, in particular, when narrow resonance structures like the $\rho$-meson lead to rapid variations in energy. In this context, pion-pion scattering has been a prime subject for lattice QCD calculations, in isospin $I=2$~\cite{Sharpe:1992pp, Kuramashi:1993ka, Gupta:1993rn, Yamazaki:2004qb, Aoki:2005uf, Beane:2005rj, Beane:2007xs, Feng:2009ij, Yagi:2011jn, Fu:2013ffa, Sasaki:2013vxa, Kawai:2017goq, Dudek:2012gj, Bulava:2016mks, Akahoshi:2019klc, Helmes:2015gla, Helmes:2019dpy, Horz:2019rrn}, $I=1$~\cite{Aoki:2007rd, Feng:2010es, Gockeler:2008kc, Lang:2011mn, Bali:2015gji, Guo:2016zos, Pelissier:2012pi, Aoki:2011yj, Dudek:2012xn, Feng:2014gba, Metivet:2014bga, Wilson:2015dqa, Alexandrou:2017mpi, Andersen:2018mau, Fu:2016itp, Werner:2019hxc}, and $I=0$~\cite{Briceno:2016mjc, Liu:2016cba, Briceno:2017qmb, Fu:2017apw, Guo:2018zss} channels. The $\pi K$ and $K\bar K$ scattering has been studied in Refs.~\cite{Beane:2006gj, Lang:2012sv, Fu:2011xw, Sasaki:2013vxa, Prelovsek:2013ela, Janowski:2014uda, Bali:2015gji, Helmes:2017smr, Brett:2018jqw, Helmes:2018nug, Wilson:2019wfr}, $\pi\phi$ and $\pi\omega$ scattering has been addressed in the recent paper~\cite{Woss:2019hse}, while $I=2$ $\pi\rho$ scattering was calculated in Ref.~\cite{Woss:2018irj}, at pion masses sufficiently large to make the $\rho$-meson stable so that Lüscher's method might be used to extract phase shifts (see also Ref.~\cite{Lang:2014tia} for a similar case).  The scattering of mesons containing heavy quarks has been considered in Refs.~\cite{Gayer:2021xzv,  Cheung:2020mql, Prelovsek:2020eiw, Piemonte:2019cbi, Bali:2017pdv, Lang:2016jpk, Albaladejo:2016lbb, Lang:2015sba, Lang:2014yfa, Torres:2014vna, Mohler:2013rwa}, see also Ref.~\cite{Guo:2018tjx}. Furthermore, despite the increased complexity of the pertinent LQCD calculations, the field of excited baryons has seen a remarkable progress~\cite{Alexandrou:2015hxa, Alexandrou:2013ata, Engel:2013ig, Dudek:2012ag, Edwards:2012fx, Briceno:2012wt, Edwards:2011jj,Bulava:2010yg, Durr:2008zz, Burch:2006cc, Alexandrou:2008tn, Menadue:2011pd, Melnitchouk:2002eg}. Meson-baryon scattering amplitudes have been calculated in Refs.~\cite{Silvi:2021uya,Stokes:2019zdd, Andersen:2017una,Lang:2016hnn, Lang:2012db}, using L\"uscher's method~\cite{Luscher:1986pf}, see also Refs.~\cite{Doring:2013glu, Wu:2016ixr}. A natural generalization of the L\"uscher approach in the two-body sector consists in the inclusion of coupled two-body channels~\cite{Liu:2005kr, Lage:2009zv, Bernard:2010fp, Doring:2011vk, Doring:2011ip, Doring:2011nd, Briceno:2012yi, Doring:2012eu, Briceno:2014oea}. The application of the method in meson-meson scattering has been spearheaded by the Hadron Spectrum Collaboration~\cite{Johnson:2020ilc, Briceno:2017qmb, Moir:2016srx, Briceno:2016mjc, Dudek:2016cru, Wilson:2015dqa, Wilson:2014cna, Dudek:2014qha} with a recent highlight given by an eight-channel analysis of an exotic $\pi_1$ meson~\cite{Woss:2020ayi}. For reviews of the two-particle coupled channel sector, see Refs.~\cite{Briceno:2017max, Detmold:2019ghl, Lang:2007mq, Doering:2014fpa, Briceno:2014tqa}. In this review we can only discuss some of these developments, see Sec.~\ref{sec:chiral-extrapolations}.

The discussion of similar finite-volume techniques for three-hadron systems is one of
the major goals of this manuscript and will be carried out in Sec.~\ref{sec:3-body-fin-vol}.
There, we review the so-called  quantization condition (an analog of the L\"uscher equation), which relates the finite-volume three-particle spectrum to the
parameters of the $S$-matrix. An alternative approach to the problem focuses on
the calculation of the interaction-induced shifts of three- and more particle energy
levels in perturbation theory~\cite{Lee:1957zzb, Huang:1957im, Wu:1959zz, Tan:2007bg, Beane:2007qr, Detmold:2008gh, Beane:2020ycc, Pang:2019dfe, Hansen:2016fzj, Romero-Lopez:2020rdq}. Yet another methodology to access multi-particle amplitudes utilizes the so-called ordered double limit~\cite{DeWitt:1956be} $(\lim_{{\rm Im}E\to 0+}\lim_{L\to \infty})$, with $E$ denoting the total energy of the system. Such an approach was used in Ref.~\cite{Agadjanov:2016mao}, extracting (complex-valued) amplitudes. For related works see Refs.~\cite{Hansen:2017mnd, Guo:2020ikh, Briceno:2020rar}. These alternatives are outside of the scope of the present review.

Furthermore, finite-volume techniques can be used to study decays of resonances,
as well as the matrix elements of currents. This will be discussed in Sec.~\ref{sec:3-LL}, where we
reflect on the latest developments related to the derivation of a three-body analog of the
Lellouch-L\"uscher formula that enables one to measure  three-body decay amplitudes
on the lattice~\cite{Muller:2020wjo,Hansen:2021ofl}.

Back to hadron spectroscopy on the lattice, we note that LQCD calculations are often carried out with quark masses larger than the physical ones. Hence, physical observables are obtained by using extrapolations in quark masses. Recently, simulations at physical quark masses in the
two- and three-particle sectors have become feasible, see,
e.g., Refs.~\cite{Bali:2015gji, Fischer:2020jzp, Fischer:2020fvl}. One might argue that, with
the advance of computing capabilities and better algorithms,  extrapolations will
soon become superfluous. However, the energy window between elastic and 
inelastic channels narrows when approaching lower pion masses, see, e.g.,
Refs.~\cite{Fischer:2020jzp, Alexandru:2020xqf}. Thus,  access to phase shifts or
other relevant quantities will be complicated by complex multiparticle dynamics and
finite-volume effects. Then, exploring QCD resonance dynamics at heavier than physical
pion masses and performing  chiral extrapolations to the physical point  represents a reasonable strategy.

The natural method to do this is to use input from Chiral Perturbation Theory (ChPT), as
discussed in Sec.~\ref{sec:chiral-extrapolations}. At its core, ChPT relies on the expansion
of the QCD Green's functions in small meson masses and momenta. Hence, the
information about the quark mass dependence is encoded there by construction. On the
other hand, the coefficients of such an expansion require input from either experiment
or lattice calculations. Thus, there is a mutually advantageous relationship between
lattice QCD and ChPT, which will help to advance our understanding of hadronic dynamics.
This is illustrated in the right panel of
Fig.~\ref{fig:fin-inf--chiral-concepts} where the
P-wave $\pi\pi$ phase shift is shown for different pion masses.
At the physical point ($M_\pi=M_\pi^{\text{ phys}}$), the chiral extrapolation should coincide
with the experiment. As the pion mass increases, the $\rho$ meson becomes narrower
and LQCD can be used to scan the energy dependence at these unphysical masses.
Unitary extensions of ChPT provide extrapolations both in the energy $\sqrt{\sigma}$ and
pion mass, as the figure demonstrates. Turning the argument around, one may use
ChPT parametrizations to fit the data both for physical and unphysical masses that allows
for the most efficient use of all available information.

In the present review, we discuss the extraction of observables from lattice QCD calculations in the two- and three-particle sectors with a focus on the merger of finite-volume approach with methods of ChPT. The layout of the paper is as follows. In Sec.~\ref{sec:chiral-extrapolations} we briefly consider foundations of unitarized ChPT and applications in the two-particle sector. Sec.~\ref{sec:3-body-fin-vol} contains a brief review of existing three-body approaches in a finite volume. Applications of these approaches for the analysis of lattice data, with a focus on chiral extrapolations, are described in Sec.~\ref{sec:applications}. Finally, latest developments, related to the treatment of three-particle decays in a finite volume, are considered in Sec.~\ref{sec:3-LL}.

\section{Chiral extrapolations}
\label{sec:chiral-extrapolations}

Chiral Perturbation Theory~\cite{Gasser:1983yg, Weinberg:1978kz} and extensions thereof to the strangeness~\cite{Gasser:1984gg} and baryon sectors~\cite{Gasser:1987rb, Bernard:1992qa, Tang:1996ca, Becher:1999he, Ellis:1997kc} is a systematic approach to low-energy QCD, which in many cases provides a benchmark for the calculations of observables in the (sub)thres\-hold energy region~\cite{Bernard:2006gx, Bernard:2007zu, Scherer:2002tk, Meissner:1993ah, Kubis:2007iy, Bernard:1995dp,Bernard:1993fp}. By construction, ChPT offers an ideal tool for extrapolations of LQCD results to the physical quark masses. See the extensive discussions in the FLAG review~\cite{Aoki:2019cca}. 

If one aims at extracting complex resonance pole positions and residues of the scattering amplitude, one has to combine ChPT with the general principles of  $S$-matrix theory: unitarity, analyticity and crossing symmetry. The list of successful applications of the method is extensive and covers both meson and baryon sectors, see, e.g., the reviews~\cite{Pelaez:2015qba, Mai:2020ltx} and references therein. In the context of chiral extrapolations on the lattice, we highlight 
pioneering works extending unitarized chiral methods to unphysical quark masses~\cite{Hanhart:2008mx, Nebreda:2010wv},
an early paper on the $\rho$ extrapolation~\cite{Bolton:2015psa}, extrapolations of the (isoscalar) $f_0(500)$~\cite{Doring:2016bdr, Guo:2018zss}, and a global analysis of $\pi\pi$ scattering in channels with different quantum numbers~\cite{Mai:2019pqr} that is discussed below in more detail. In most of these approaches, actual lattice data were analyzed, but chiral extrapolations have also been performed for other hadronic reactions~\cite{Niehus:2019nkl, Dax:2018rvs}. Chiral extrapolations have been implemented in recent three-body studies as well, addressing the two-body subsystem at unphysical pion masses. These developments will be briefly discussed in Sec.~\ref{subsec:chiral-3body}.

\begin{figure}
    \centering
    \includegraphics[width=0.95\linewidth,trim=0 5cm 0 4.cm,clip]{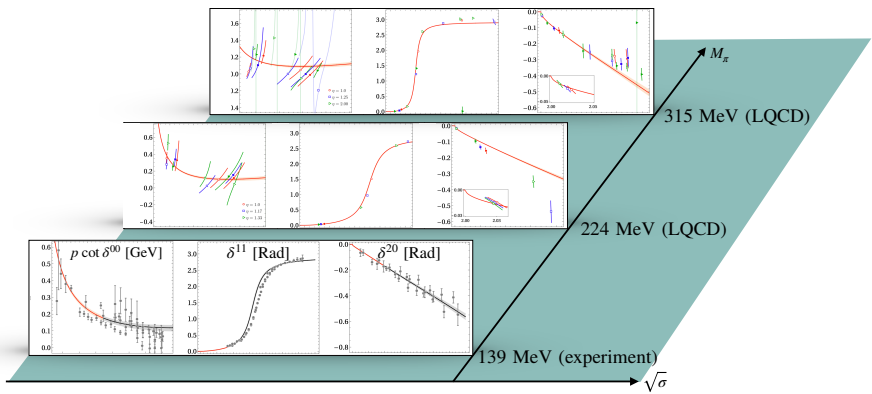}
    \caption{
      Phase shifts of two-pion scattering in all isospin channels at physical
      and unphysical pion masses. Red curves show the result of the mIAM
      global fits~\cite{Mai:2019pqr} to the GWQCD lattice
      data~\cite{Guo:2018zss, Guo:2016zos, Culver:2019qtx} at
      $M_\pi\simeq 224$~MeV and $M_\pi\simeq 315$~MeV. Experimental
      results~\cite{Batley:2010zza, Froggatt:1977hu, Estabrooks:1974vu, Hyams:1973zf, Protopopescu:1973sh, Grayer:1974cr, Rosselet:1976pu, Janssen:1994wn, Estabrooks:1974vu} (gray data points) are not part of the fit and are plotted for comparison only.
    }
    \label{fig:simultaneous-pipi-GWUQCD}
\end{figure}

There are numerous lattice calculations in the two-pion sector, see e.g., Refs.~\cite{Guo:2018zss, Guo:2016zos, Culver:2019qtx, Wilson:2015dqa, Wilson:2014cna, Fischer:2020jzp}. These calculations cover a wide energy range and involve pions with the masses varying from the physical value up to $400$~MeV. To make a full use of the available information {\em at different pion masses,} a parametrization of the scattering amplitude is required, which not only exhibits the correct analytic properties, but also implements correct chiral behavior. 

In order to achieve this goal, various approaches have been used in the past. For example, the chiral unitary approach in SU(3) with NLO contact terms~\cite{Oller:1998hw}, which has been applied in Refs.~\cite{Guo:2016zos, Hu:2016shf, Hu:2017wli} for extrapolations of the $\rho$ resonance (see also Refs.~\cite{Oller:1997ti, Albaladejo:2008qa} for related approaches). Unitarized $U(3)$ ChPT~\cite{Guo:2011pa, Guo:2012ym, Guo:2015xva} has been used for the $\pi\eta^{(\prime)}-K\bar K$ coupled-channel extrapolation of the $a_0(980)$~\cite{Guo:2016zep}.

Another possibility for amplitude construction is provided by the so-called modified Inverse Amplitude Method (mIAM)~\cite{Truong:1988zp, Dobado:1996ps, GomezNicola:2007qj, Hanhart:2008mx}. Besides the usual constraints from perturbative ChPT at a given order, mIAM also fulfils chiral constraints on resonance trajectories~\cite{Bruns:2017gix}. While implementing the elastic unitarity exactly, this method relies on the full next-to-leading (NLO) chiral scattering amplitudes, obeying crossing symmetry at this order. The mIAM amplitude is independent of the renormalization scale, occurring in  the UV-divergent loops. Note that extensions to two loops have also been worked out~\cite{Pelaez:2006nj}, including a chiral extrapolation of the $\rho$ resonance~\cite{Niehus:2020gmf}, see Ref.~\cite{Pelaez:2021dak} for a recent review.

Technically, mIAM is based on the leading-order ($T_2^{I\ell}(s)$) and the NLO ($T_4^{I\ell}(s)$) chiral amplitudes, projected onto a given isospin $I$ and angular momentum $\ell$. A unitary scattering amplitude $T_{\rm mIAM}^{I\ell}(s)$ can then be derived~\cite{Truong:1988zp}, using dispersion relations:
\begin{align}
T_{\rm mIAM}^{I\ell}(s)&=\frac{(T_2^{I\ell}(s))^2}{T_2^{I\ell}(s)-T_4^{I\ell}(s)+A^{I\ell}_m(s)}\, ,\nonumber\\
A^{I\ell}_m(s)&=T^{I\ell}_4(s_2)
-\frac{(s_2-s_A)(s-s_2)}{s-s_A}\left(\frac{\partial T^{I\ell}_2}{\partial s}(s_2)-\frac{\partial T^{I\ell}_4}{\partial s}(s_2)\right)\,.
\label{eq:mIAM}
\end{align}
The term $A^{I\ell}_m(s)$ has been introduced in Refs.~\cite{FernandezFraile:2007fv,GomezNicola:2007qj}, in order to avoid the
appearance of an unphysical pole at $T_2(s)=T_4(s)$. 
Further, $s_A$ denotes the position of the Adler zero at NLO, given by the equation $T_2^{I\ell}(s_A)+T_4^{I\ell}(s_A)=0$,
and $s_2$ stands for the same quantity at leading order, obeying the equation $T_2^{I\ell}(s_2)=0$.

The leading order chiral amplitude is a function of energy, Goldstone-boson mass, $M^2=B(m_u+m_d)$, and pion decay constant in the chiral limit, $f_0$. In the two-flavor case, the amplitude $T_4^{I\ell}$ involves two low-energy constants (LECs) $\bar l_1$ and $\bar l_2$. Two additional low-energy constants $\bar l_3$, $\bar l_4$ enter the NLO chiral amplitude,  when $M,f_0$ are replaced by the physical pion mass and pion decay
constant, using one-loop results~\cite{Gasser:1983yg}:
\begin{align}
M_\pi^2=M^2\left(1-\frac{M^2}{32\pi^2f_0^2}\bar l_3\right)
\,,\qquad
f_\pi=f_0\left(1+\frac{M^2}{16\pi^2f_0^2}\bar l_4\right)\,.
\end{align}

For the description of the finite-volume LQCD spectra, various strategies have been applied in the past~\cite{Doring:2016bdr,Guo:2018zss,Molina:2020qpw,Fischer:2020jzp}. The most natural approach (see Ref.~\cite{Fischer:2020fvl,Mai:2019pqr,Brett:2021wyd}) is to re-write
Eq.~\eqref{eq:mIAM} in terms of the $K$-matrix, or 
\begin{equation}
  \label{eq:IAMphase}
  \cot\,\delta_{I\ell}^\mathrm{mIAM}(s)=
  \frac{\sqrt{s}}{2k}
  \Bigg(\frac{T_2^{I\ell}(s)-\bar T_4^{I\ell}(s)+A_m^{I\ell}(s)}{(T_2(s))^2}
  -16\pi\operatorname{Re}{J(s)}\Bigg)\,,
\end{equation}
where $k$ is the magnitude of the center-of-mass meson momentum. Further, $\bar T^{I\ell}_4=T_4^{I\ell}(s)-T_2^{I\ell}(s)J(s)T_2^{I\ell}(s)$, where $J(s)$ denotes the meson-meson loop in dimensional regularization. The finite-volume energy spectrum is determined by the L\"uscher equation, with the phase shift given by Eq.~(\ref{eq:IAMphase})~\cite{Luscher:1986pf, Luscher:1990ux, Luscher:1990ck, Werner:2019hxc, Bulava:2016mks, Morningstar:2017spu}. The two-body parametrization with mIAM in infinite and finite volume is also discussed in detail in the Appendix of Ref.~\cite{Brett:2021wyd}. See Ref.~\cite{Albaladejo:2012jr} for estimates of exponentially suppressed contributions for IAM.

As already mentioned, mIAM was applied in different isospin channels of $\pi\pi$ scattering~\cite{Doring:2016bdr, Guo:2018zss, Mai:2019pqr, Fischer:2020jzp}, leading to a very consistent picture across all three channels. For example, the result of a simultaneous analysis of finite-volume spectra, obtained by the GWUQCD collaboration~\cite{Guo:2018zss, Guo:2016zos, Culver:2019qtx}, including also correlations between different isospin channels, is depicted in Fig.~\ref{fig:simultaneous-pipi-GWUQCD}. The application of mIAM to the three-flavor sector is straightforward, but more tedious due to the fact that the strange quark mass is, typically, treated differently from that of the light ones. One way~\cite{Andersen:2018mau} is to fix the sum of all three quarks to the physical value. Alternatively, one could assume that the strange quark mass is fixed to its physical value~\cite{Wilson:2015dqa, Dudek:2012xn, Bulava:2016mks, Feng:2014gba, Alexandrou:2017mpi}. LQCD results for the isovector channel along these two $m_s(m_l)$ ``trajectories'' have been studied recently in Ref.~\cite{Molina:2020qpw}, using Inverse Amplitude Method, which is identical to the  mIAM for this channel. Note that  $SU(3)$ ChPT to one loop in the IAM implementation was recently used in the first ever
extrapolation for a system of three kaons at maximal isospin~\cite{Alexandru:2020xqf}, see  Sec.~\ref{sec:threek}.

Finally, we wish to address the scale setting issue in the analysis of the LQCD results. First, we note that all involved scattering amplitudes are expressed in powers of a dimensionless quantity $M^2/(4\pi f_0)^2$. Thus one can hope to perform an analysis directly in lattice units. Still, the scattering amplitudes beyond leading order depend explicitly on the LECs $\bar l_i$, which are independent of the renormalization scale $\mu$, but depend on the quark masses. In order to perform the extrapolation in the quark masses, it is necessary to introduce the quark-mass independent renormalized LECs, which then depend on $\mu$:
\begin{equation}
\label{eq:LECS}
l_i^r=\frac{\gamma_i}{32\pi^2}\left(\bar l_i+\log \frac{M^2}{\mu^2}\right)\, ,\quad 
\quad
\gamma_1=\frac{1}{3},\,\gamma_2=\frac{2}{3},\,\gamma_3=-\frac{1}{2},\,\gamma_4=2\,.
\end{equation}
Hence, for a fixed scale $\mu$, one can make predictions for two-particle scattering at a different pion mass. Of course, fixing the $\mu$ to some dimensionfull value requires to perform the scale setting. This issue was discussed at length in Ref.~\cite{Mai:2019pqr}, where the influence of the $\log{\mu}$ term in Eq.~\eqref{eq:LECS} was found to be well below the statistical uncertainties. Additionally, in some specific cases, such as the isovector channel, the chiral NLO amplitude $T_4^{11}$ depends only on the difference $\bar l_{12}=\bar l_1-\bar l_2$, which leads to an exact cancelling of the $\log{\mu}$ term and allows one to perform the analysis entirely in lattice units~\cite{Fischer:2020jzp}.

\section{Three-body Quantization Condition}
\label{sec:3-body-fin-vol}

Recent years have witnessed a rapid increase of interest to the investigation of three-particle dynamics from lattice calculations. Such challenging studies became feasible only lately, owing to the increased computational resources, as well as the progress achieved on methods and algorithms~\cite{Beane:2007es, Detmold:2008fn, Detmold:2008yn, Blanton:2019vdk, Horz:2019rrn, Culver:2019vvu, Fischer:2020jzp, Hansen:2020otl, Alexandru:2020xqf, Romero-Lopez:2018rcb, Romero-Lopez:2020rdq}. In its turn, this progress was triggered by the development of the formalism that allows the mapping of the three-particle spectrum on the scattering observables in the two- and three-particle sectors, the so-called {\em three-particle quantization condition}. The work in this direction started in 2012~\cite{Polejaeva:2012ut}. During the next few years, the three-body quantization condition has been derived in three different but equivalent 
frameworks, usually termed as Relativistic Field Theory (RFT)~\cite{Hansen:2014eka, Hansen:2015zga}, the Non-Relativistic Effective Field Theory (NREFT)~\cite{Hammer:2017uqm, Hammer:2017kms} and Finite Volume Unitarity (FVU)~\cite{Mai:2017bge, Mai:2018djl} approaches, see Refs.~\cite{Hansen:2019nir, Rusetsky:2019gyk} for recent reviews. Recently, another approach based on time-ordered perturbation theory was derived~\cite{Blanton:2020gha}, and used to relate the above approaches~\cite{Blanton:2020jnm}.  We note also the earlier work~\cite{Briceno:2012rv}, where the three-body analog of the L\"uscher equation has been written down in the particle-dimer picture. Overall, this development has boosted activities in the field, as seen, e.g., from Refs.~\cite{Roca:2012rx, Bour:2012hn, Meissner:2014dea, Jansen:2015lha, Hansen:2015zta, Hansen:2016fzj, Guo:2016fgl, Konig:2017krd, Briceno:2017tce, Sharpe:2017jej, Guo:2017crd, Guo:2017ism, Meng:2017jgx, Guo:2018ibd, Guo:2018xbv, Klos:2018sen, Briceno:2018mlh, Briceno:2018aml, Doring:2018xxx, Jackura:2019bmu, Mai:2019fba, Guo:2019hih, Blanton:2019igq, Briceno:2019muc, Romero-Lopez:2019qrt, Pang:2019dfe, Guo:2019ogp, Zhu:2019dho, Pang:2020pkl, Hansen:2020zhy, Guo:2020spn, Guo:2020wbl,Konig:2020lzo, Guo:2020ikh, Blanton:2020gha, Blanton:2020gmf, Muller:2020vtt, Brett:2021wyd}. Note also earlier work on related issues~\cite{Kreuzer:2010ti, Kreuzer:2009jp, Kreuzer:2008bi, Kreuzer:2012sr}.

Three alternative but essentially equivalent versions of the three-body quantization condition are highlighted below. Those are the aforementioned NREFT~\cite{Hammer:2017uqm, Hammer:2017kms}, RFT~\cite{Hansen:2014eka, Hansen:2015zga} and FVU~\cite{Mai:2017bge,Mai:2018djl} approaches. Their applications in the analysis of lattice data will be reviewed in Sec.~\ref{sec:applications}. In the following, we will consider the main features of three formalisms and the links between them. Obviously, the details are too lengthy for this short review, and  we refer the reader to the corresponding original papers. To ease the notation, we additionally restrict the discussion to the case of three identical particles of a mass $m$. In addition, $Z^2$ symmetry is assumed to hold, under which all Green's functions with an odd number of external legs vanish identically.

The common feature of all approaches is the identification of the appearing intermediate states in the scattering amplitude that can go on-shell. Such intermediate states are the only source of the power-law (in $L$) finite-volume corrections, while the off-shell states give only exponentially suppressed corrections in $L$, see, e.g., Ref.~\cite{Luscher:1990ux}. Taking $L$ much larger than the inverse mass of a particle, one may neglect these exponentially suppressed corrections completely. Restricting the center-of-mass energy of three particles to $3m<\sqrt{s}<5m$, only the three-body intermediate states can go on-shell. Then, in order to arrive at the quantization condition, one has to re-sum all such contributions in a finite volume and search for the position of the poles of the Green's function. At this point, each of the above-mentioned formalisms takes a different path:
\begin{itemize}

\item[$\blacksquare$]
The NREFT formalism~\cite{Hammer:2017uqm, Hammer:2017kms} is based on non-relativistic effective Lagrangians. Thus, only the forward propagation in time is allowed, and the virtual creation/an\-ni\-hi\-la\-ti\-on into pairs is prohibited but is implicitly contained in the effective couplings. Within this setting, only three-body intermediate states emerge by construction, when the total energy in the center-of-mass frame is below $5m$. In addition, the approach of Refs.~\cite{Hammer:2017uqm, Hammer:2017kms} uses the so-called particle-dimer picture that renders the bookkeeping of Feynman diagrams in the three-particle case very simple and transparent. It should be stressed that the particle-dimer picture is not an approximation but a mathematically equivalent description of a three-particle system. Also, it does not imply the existence of a stable two-body bound state (the same remarks apply to isobars used in the FVU approach discussed below).\smallskip

The three-particle quantization condition follows from the Faddeev equation for the particle-dimer scattering amplitude, written down in a finite volume, where all three-momenta are discretized as $\bm{p}=2\pi\bm{n}/L$ with $\bm{n}\in\mathds{Z}^3$. Such an amplitude is singular at the energies corresponding to the eigenvalues of the Hamiltonian in a finite volume. In the simplest case, assuming that pair interactions occur only in the S-wave and the particle-dimer interaction contains only the non-derivative term with  coupling $H_0$, the quantization condition takes the form:
\begin{align}
\label{eq:NREFT}
0&=\det\left(\hat\tau_L(E)^{-1}-Z(E)\right)\,,
\\[2mm]\nonumber
[Z(E)]_{\bm{p}\bm{q}}&=
\frac{1}{\bm{p}^2+\bm{q}^2+\bm{p}\bm{q}-mE}+\frac{H_0(\Lambda)}{\Lambda^2}\,,
\\[2mm]\nonumber
8\pi[\hat\tau_L(E)]^{-1}_{\bm{p}\bm{q}}&=
\delta_{\bm{p}\bm{q}}
\left(p^*\cot\delta(p^*)-\frac{4\pi}{L^3}\sum_{\bm{l}}\frac{1}{\bm{p}^2+\bm{l}^2+\bm{p}\bm{l}-mE}
\right)\,.
\end{align}
Here $E=\sqrt{s}-3m$, whereas $Z$ and $\hat\tau_L$ are the three-body kernel of the Faddeev equation and the finite-volume two-body scattering amplitude, respectively. Each of these elements enters as a matrix in the space of spectator momenta. We also note here that the UV behaviour of the spectator momenta is regulated by a hard cutoff $\Lambda$, while in the sum, entering the definition of $\hat\tau_L^{-1}$, the dimensional regularization is implicit. The S-wave phase shift is denoted by $\delta(p^*)$ for $p^*=\sqrt{3/4\,\bm{p}^2 -mE}$.\smallskip\smallskip

Note that, although the above quantization condition was derived, dropping higher
partial waves in pair interactions, as well as higher order (derivative) particle-dimer
couplings, these can be taken into account without much ado.
Relativistic corrections can also be included systematically. It should be realized that
the crucial feature of NREFT is barring the backward propagation as well as
the pair creation/annihilation, and not the use of the non-relativistic dispersion law for a
single particle. In fact, the relativistic kinematics in this approach can be easily implemented
along the lines suggested in Refs.~\cite{Colangelo:2006va,Gasser:2011ju}. At the leading order this was demonstrated in Ref.~\cite{Muller:2020wjo}, and the generalization to higher orders is in progress.

\item[$\blacksquare$]
The RFT formalism~\cite{Hansen:2014eka, Hansen:2015zga} approaches the problem using relativistic Feynman diagrams. As a result, one has to sort all emerging diagrams into  reducible ones (i.e., those which can be made disconnected by cutting exactly three particle lines) and the irreducible ones (all the rest). The irreducible diagrams are replaced by their infinite-volume counterparts, dropping  exponentially suppressed contributions. Collecting all terms that lead to power-law corrections, the RFT quantization condition in the simplest case takes the form: 
\begin{align}
\label{eq:RFT}
0&=\det\left(
L^3\left(\tilde F/3-\tilde F(\tilde K_2^{-1}+\tilde F+\tilde G)^{-1}\tilde F\right)^{-1}+K_{\rm df,3}
\right)\,,
\\[3pt]\nonumber
[\tilde F(\sqrt{s})]_{\bm{p}\bm{q}}&=\delta_{\bm{p}\bm{q}}\frac{H(\bm{p})}{4E_{\bm{p}}}
\left(
\frac{1}{L^3}\sum_{\bm{a}}-\int_{\rm PV} \frac{d^3a}{(2\pi)^3}
\right)
\frac{H_2(\bm{a},-\bm{p}-\bm{q})}{4E_{\bm{a}}E_{\bm{p+q}}(\sqrt{s}-E_{\bm{p}}-E_{\bm{a}}-E_{\bm{p}+\bm{q}})}\,,\\[2pt]\nonumber
[\tilde G(\sqrt{s})]_{\bm{p}\bm{q}}&=\frac{H(\bm{p})H(\bm{q})}{L^34E_{\bm{p}}E_{\bm{q}}((\sqrt{s}-E_{\bm{p}}-E_{\bm{q}})^2-(\bm{p}+\bm{q})^2-m^2)}
\,,\\[2pt]\nonumber
[\tilde K_2(\sqrt{s})]_{\bm{p}\bm{q}}&=\delta_{\bm{p}\bm{q}}
\frac{32\pi E_{\bm{p}}\sqrt{\sigma_p}}{p^*\cot\delta(p^*)+|p^*|(1-H(\bm{p}))}\,.
\end{align}
Further details of the derivation can be found in a comprehensive summary given in Ref.~\cite{Hansen:2019nir}. Again, the quantization condition arises as a determinant equation with respect to the spectator momenta. In that, the functions $H(\bm{p})$, $H_2(\bm{p})$ provide a smooth cutoff at large values of momenta. Additionally, the kinematical variables are defined as $\sigma_x=(\sqrt{s}-E_{\bm{x}})^2-\bm{x}^2$ and $ E_{\bm{x}}=\sqrt{\bm{x}^2+m^2}$, while $p^*$ stands for the magnitude of the relative three-momenta defined in the two-body subsystem. The two-body interactions are encoded in the $K$-matrix related quantity $\tilde K_2$, while $K_{\rm df,3}$ parametrizes the genuine three-body force. Generalizations of Eq.~\eqref{eq:RFT} to two-particle subsystems with spin, non-identical particles, and moving frames, etc., have been worked out in Refs.~\cite{Hansen:2020zhy,Blanton:2019igq}.\smallskip

Note that the propagator $\tilde G$ that describes the rearrangement between the particle pair and the spectator, is written down in relativistically-invariant form. This on-shell form does not emerge from the beginning in perturbation theory. However, since the difference between $\tilde G$ and the original expression obtained in perturbation theory represents a low-energy polynomial, it was possible to rewrite the quantization equation in terms of the $\tilde G$~\cite{Briceno:2018aml} -- the difference should be accounted for by the change in the three-body regular term $K_{\rm df,3}$. The choice of the relativistic-invariant form for $\tilde G$ has far-reaching implications. It can be shown now that the $K_{\rm df,3}$ in this equation has to be relativistic-invariant as well. Moreover, the quantization condition in non-rest frames can be straightforwardly written down in terms of the same $K_{\rm df,3}$.

\item[$\blacksquare$] The FVU formalism~\cite{Mai:2017bge,Mai:2018djl} takes another approach, starting from three-body unitarity as a guiding principle for the construction of a relativistic three-body scattering amplitude. This approach is based on the observation that the only diagrams, which lead to the power-law corrections in a finite volume, are those that contribute to three-body unitarity~through their imaginary parts~\cite{Aaron:1969my, Mai:2017vot}, see also Refs.~\cite{Jackura:2018xnx, Jackura:2020bsk, Dawid:2020uhn}. In order to carry out practical implementations of this idea, the dynamics of the three-body system in the FVU approach is formally separated into a cluster of two-body states (called isobar) and a spectator. This allows one to derive analytic constraints on the form of the isobar propagator as well as the Bethe-Salpeter kernel for the isobar-spectator interaction\footnote{Note that a relativistic infinite-volume amplitude~\cite{Mai:2017vot}, constructed along the same lines, was used recently~\cite{Sadasivan:2020syi} to address Dalitz plots of the reaction $\tau\to\nu_\tau(a_1(1260)\to\pi\pi\pi)$.}. In a finite volume, discretizing three-momenta, this formulation leads to a three-body quantization condition. In its simplest form, this condition reads
\begin{align}
\label{eq:FVU}
0&=\det\left(B_0+C_0-E_L\left(K^{-1}/(32\pi)+\Sigma_{L}\right)
\right)\,,\\[4pt]\nonumber
[\Sigma_{L}(\sqrt{s})]_{\bm{p}\bm{q}}
&=\delta^{3}_{\bm{p}\bm{q}}\frac{\sigma_{p}}{L^3}
\sum_{{\bm{k}}
}
\frac{\sqrt{\sigma_p}}{\sqrt{s}-E_{\bm{p}}}
\frac{1}{8E_{\bm{k}^*}^3(\sigma_{p}-4E_{\bm{k}^*}^2)}
\,,\\[4pt]\nonumber
[B_0(\sqrt{s})]_{\bm{p}\bm{q}}^{-1}&=-2E_{\bm{p}+\bm{q}}(\sqrt{s}-E_{\bm{p}}-E_{\bm{q}}-E_{\bm{p}+\bm{q}})
\ ,\quad\quad [E_{L}]_{\bm{p}\bm{q}}=\delta^{3}_{\bm{p}\bm{q}}2L^3E_{\bm{p}},\\[4pt]\nonumber
[K^{-1}(s)]_{\bm{p}\bm{q}}&=\delta^{3}_{\bm{p}\bm{q}}\,p^*\cot\delta(p^*)\,.
\end{align}
The asterisk as a superscript means that the pertinent quantity is
calculated in the two-body rest frame. The UV divergences in this approach, like other approaches, can be tamed in different ways. For example, in Ref.~\cite{Mai:2018djl}, this has been achieved using smooth cutoff functions. In a more recent work~\cite{Brett:2021wyd}, the divergences of the isobar self-energy were regularized, using subtractions in $\sigma$ and putting a hard cutoff on the spectator momenta ($\bm{p}, \bm{q}$). The genuine two and three-body forces are encoded in the scattering phase shift $\delta(p^*)$ and the isobar-spectator contact term $C_0$, which is a real-valued function of energy and spectator momenta.\smallskip

The above approach makes no direct connection to a Lagrangian formalism. However, one can interpret its building blocks in a diagrammatic language as illustrated to the top of Fig.~\ref{fig:schematic}.  The term $B_0$ can be referred to as the one-particle exchange term in forward time propagation (Fig.~\ref{fig:schematic}b)
and the term $(K^{-1}/(32\pi)+\Sigma_{L})^{-1}$ of Eq.~\eqref{eq:FVU} corresponds to the isobar-spectator propagation shown in Fig.~\ref{fig:schematic}a. In addition, restoring relativistic invariance in the term $B_0$ does not violate three-body unitarity and can be accomplished, as shown explicitly in Refs.~\cite{Mai:2017vot, Sadasivan:2020syi}. Finally, we note that generalizations of the above condition to moving frames and three-flavor sector can be found in Refs.~\cite{Brett:2021wyd, Culver:2019qtx, Mai:2019fba}.

\end{itemize}

To summarize, all approaches lead to a very similar general form of the quantization condition. In contrast to the two-body case~\cite{Luscher:1984xn}, the determinant that appears in the quantization condition of Eqs.~\eqref{eq:NREFT}, \eqref{eq:RFT} or \eqref{eq:FVU} depends, in addition, on the spectator momenta. Explicit equivalence between RFT and FVU formalisms was shown in Ref.~\cite{Blanton:2020jnm}, whereas the correspondence of the infinite-volume approaches was discussed in Ref.~\cite{Jackura:2019bmu}. Also, it was shown in Ref.~\cite{Hammer:2017kms} that topologies arising in the RFT formalism have exact correspondence to those in the NREFT approach.

Finally, we briefly consider the workflow in the process of fitting the data. In case of the two-body L\"uscher equation, an observable quantity --the phase shift-- can be directly extracted from the measured energy levels. This is no longer the case in three-particle systems, where the unphysical quantities --the regular amplitudes ($H_0$, $K_{\rm df,3}$ or $C_0$, respectively)-- are determined in a first step by fitting them to the three-body lattice spectrum. Or, more generally, these quantities are fit together with the $K$-matrix to the two and three-body lattice spectrum simultaneously. In a second step, the physical amplitudes are then determined through the solution of integral equations in the infinite volume, using this input. The need of the two-step approach constitutes a major difference to the two-body case but does not pose a conceptual problem.

A final remark concerns the ``three-body forces'' $H_0$, $K_{\rm df,3}$ or $C_0$ in the different approaches. As discussed, their values cannot be compared directly because they are defined differently. In addition, they depend on the regulator of the three-body equation, as well as on the details of the chosen parametrization of the two-body input in the subthreshold region. In that sense, the three-body force is not an observable.

\section{Analysis of lattice data}
\label{sec:applications}

In this section, we review the use of the discussed finite-volume frameworks for the analysis of lattice data. Due to the limited scope of the present review, we focus on applications that involve chiral extrapolations including an outlook for these methods given in Sec.~\ref{sec:concl}.

\begin{figure}
    \centering
    \includegraphics[width=1.\linewidth]{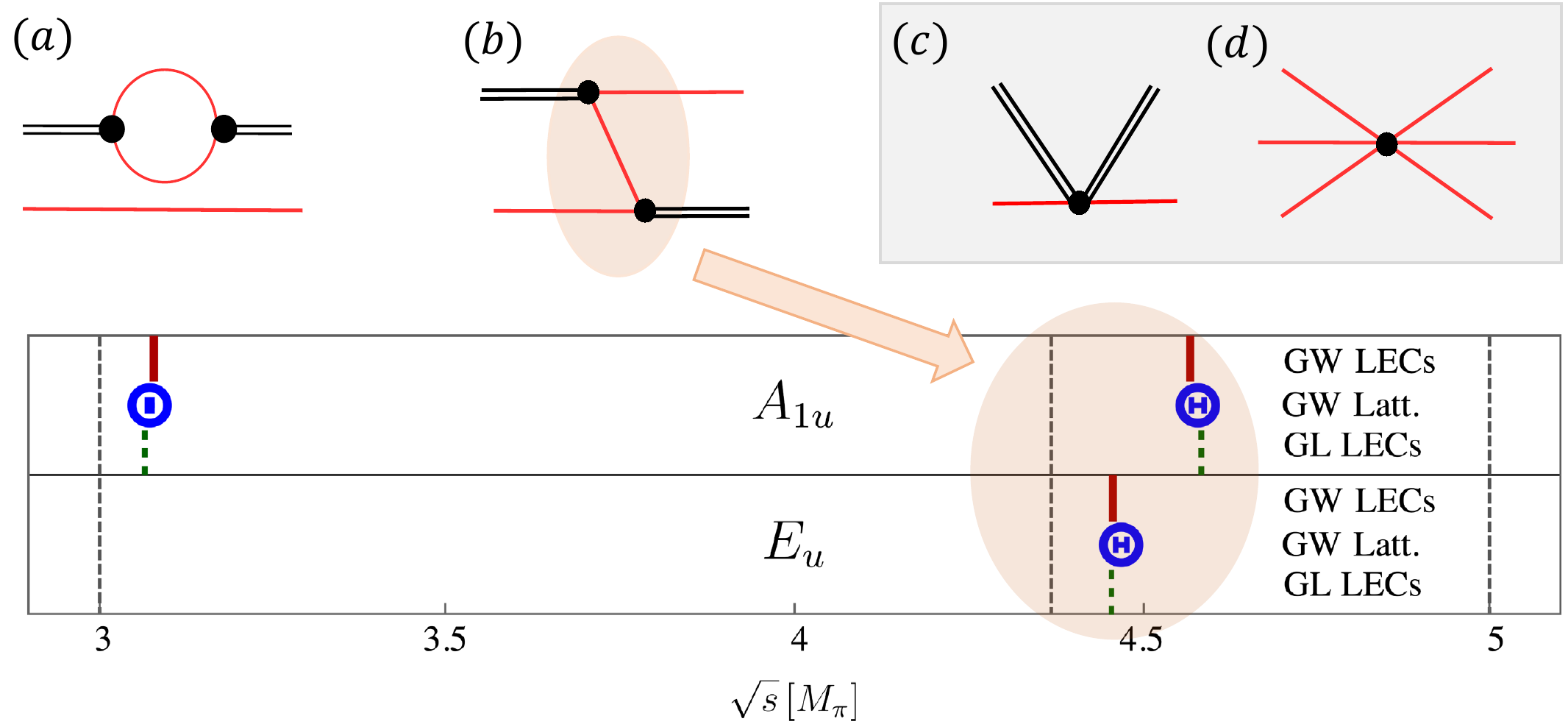}
    \caption{
{\bf Top:}
(a) Isobar-spectator propagator; (b) exchange process; (c) isobar-spectator interaction vs (d) the three-pion vertex in ChPT. 
{\bf Bottom:} 
The energy spectrum of three pions at maximal isospin from Ref.~\cite{Culver:2019vvu},
as calculated in lattice QCD (GW Latt.) and predicted from FVU with IAM
extrapolation of the two-body input using LECs from Ref.~\cite{Mai:2019pqr}
(GW, red vertical lines) and Ref.~\cite{Gasser:1983yg} (GL, dotted vertical lines).
The energy shifts from the noninteracting levels (dashed lines) are
predicted, using the interaction kernel (b) fixed from unitarity, as highlighted.
See text for further explanations.
}
\label{fig:schematic}
\end{figure}

\subsection{Chiral extrapolations for Three Pions at Maximal Isospin}
\label{subsec:chiral-3body}

In a three-pion system, ChPT describes the quark mass dependence of both the two-particle and three-particle interactions, which are encoded in the non-singular kernel ($H_0$, $K_{\rm df,3}$ or $C_0$, in different approaches). The former is far more important for non-resonant systems like three $\pi^+$. This can be
seen, e.g., from the perturbative expression of the ground-state energy shift, given in Ref.~\cite{Beane:2007qr} -- there, the contribution of the three-body force comes at next-to-next-to-next-to-leading (N$^3$LO) order in the expansion in $1/L$.

Furthermore, owing to three-particle unitarity~\cite{Mai:2017vot}, the exchange diagrams, shown in Fig.~\ref{fig:schematic}b, are determined by the same two-body input as the ones shown in Fig.~\ref{fig:schematic}a. The exchange term (in the $u$-channel) produces known relative and absolute interaction strengths in different partial waves/ir\-re\-du\-cib\-le representations (irreps) in the infinite/finite volume. Therefore, the isobar-spectator interaction in, e.g., the $A_{1u}$ and $E_u$ irreps is a prediction,
directly stemming from three-body unitarity. This prediction manifests itself in the size of the energy shifts in the excited states, as highlighted in orange in Fig.~\ref{fig:schematic}. As the figure shows, the predictions based on unitarity and lattice
data for these two irreps indeed agree, if the three-body force is set to zero; in conclusion, three-body unitarity is directly visible in the lattice data. Note that all approaches discussed in Sec.~\ref{sec:3-body-fin-vol}
contain the exchange contribution.

\begin{figure}
\begin{minipage}{0.6\linewidth}
    \includegraphics[width=\linewidth]{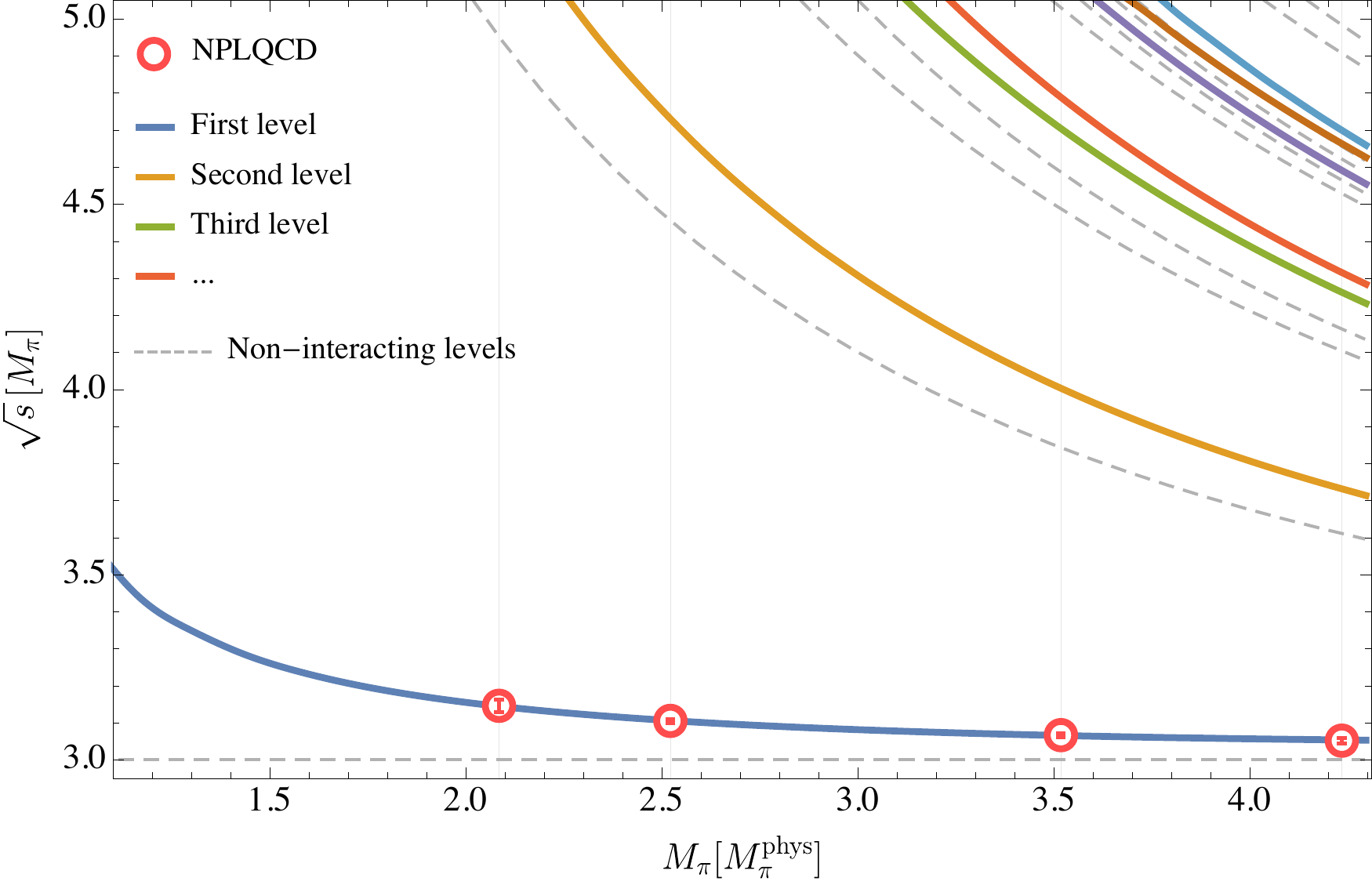}
\end{minipage}
~
\begin{minipage}{0.37\linewidth}
    \includegraphics[width=1.\linewidth,trim=0 1.4cm 0 0,clip]{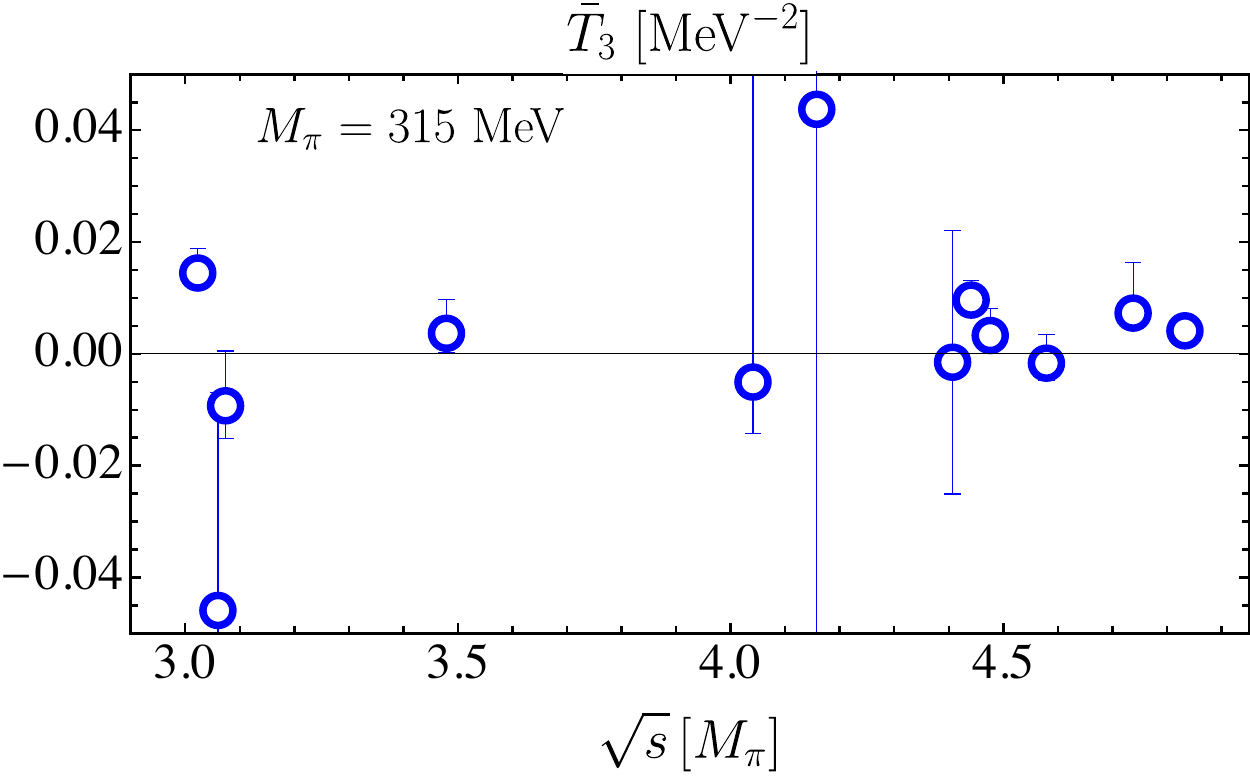}\\
    \includegraphics[width=1.\linewidth,trim=0 0 0 0.75cm,clip]{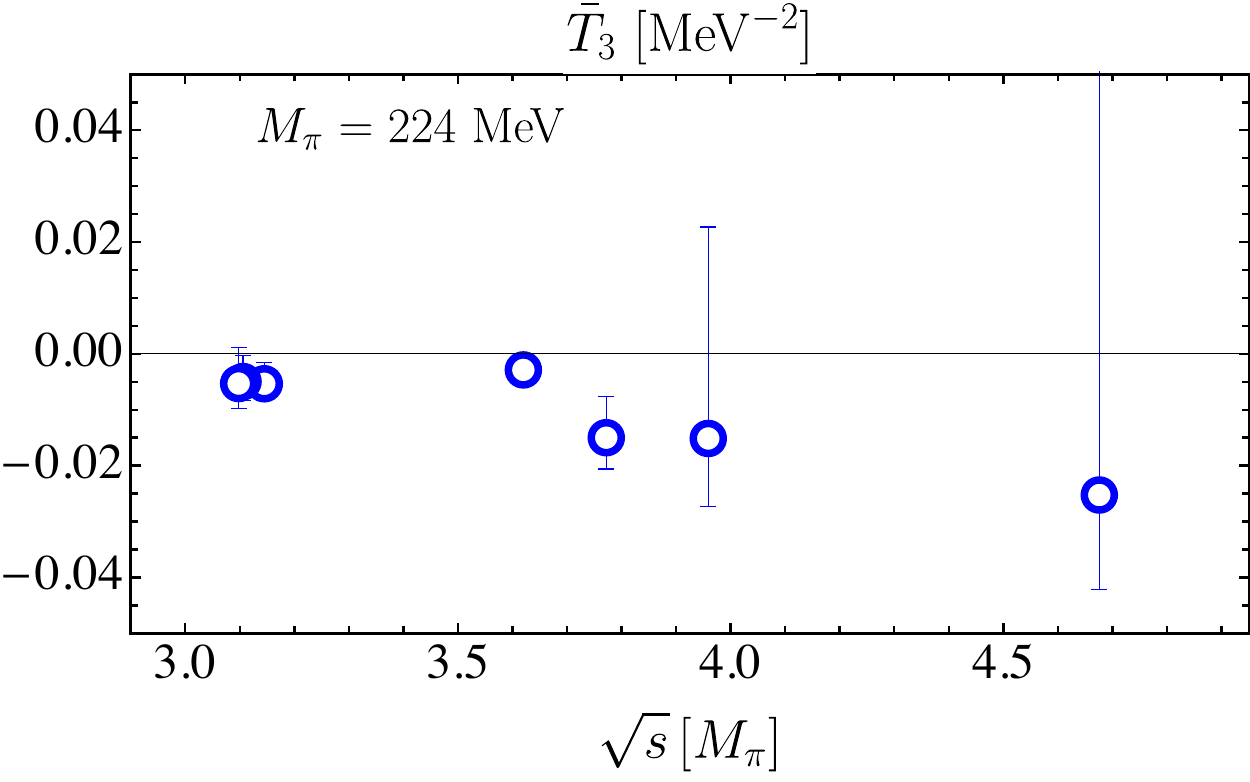}
\end{minipage}
    \caption{
      {\bf Left:} Chiral extrapolation of the $\pi^+\pi^+\pi^+$ finite-volume spectrum, taken
      from Ref.~\cite{Mai:2018djl}. Full and dashed lines denote the interacting and
      non-interacting energy eigenvalues as a function of pion mass for a fixed volume
      $L=2.5$~fm. The data in red show  results of the NPLQCD lattice calculation~\cite{Detmold:2008fn}.
      {\bf Right:} Three-body force $\bar T_3$ for two different pion masses, determined in a fit to the  single energy eigenvalues. Data are taken from Ref.~\cite{Brett:2021wyd}.
    }
    \label{fig:chiralextrapolation-3b}
\end{figure}

The first chiral extrapolation of three-body finite-volume spectra was performed in Ref.~\cite{Mai:2018djl}, using the FVU framework and IAM for the two-body input.  The main result of the study of the three-pion lattice spectrum in this approach is depicted in Fig.~\ref{fig:chiralextrapolation-3b}, left panel, showing a {\em prediction} for the excited level spectrum of three pions at maximal isospin, as a function of the pion mass for a fixed lattice volume ($L=2.5$~fm). The figure also demonstrates that
working at unphysically large pion masses can actually be an advantage, because more energy eigenvalues can be found in the elastic region $3M_\pi<\sqrt{s}<5M_\pi$, allowing for a finer-grained sampling of the amplitude. As the figure shows, the chiral extrapolation matches the NPLQCD lattice calculation for the threshold energy level. The extraction of the three-body force, also performed in Ref.~\cite{Mai:2018djl}, is discussed in the following section. 

Next, we consider the chiral extrapolation of the three-body force. Note  that since FVU does not operate directly with Feynman diagrams, a certain effort is needed to properly ``translate'' ChPT into the FVU input -- for example, the six-pion contact term in ChPT (Fig.~~\ref{fig:schematic}d) should be related to the isobar-spectator vertex, see Fig.~\ref{fig:schematic}c. This has been discussed in Ref.~\cite{Brett:2021wyd} and, in a broader context, in Ref.~\cite{Jackura:2019bmu}.

The first lattice QCD calculation producing excited states for the three-pion system at
maximal isospin was performed in Ref.~\cite{Horz:2019rrn}. Using, again, ChPT to NLO
for the two-body input, and extending the FVU formalism to moving frames and different
irreps (the latter based on Ref.~\cite{Doring:2018xxx}), these levels were predicted in
Ref.~\cite{Mai:2019fba}. Assuming a vanishing three-body force resulted in a
$\chi^2_\text{dof}\approx 0.86$ for the three-body sector and
$\chi^2_\text{dof}\approx 1.79$ for the combined (correlated) two- and three-body
sectors.
We also note here that the measurements of the ground-state and excited
two- and three-pion levels, carried out in Ref.~\cite{Horz:2019rrn}, follow the
perturbative predictions of the NREFT approach~\cite{Pang:2019dfe}.

Finally, it should be mentioned that the three-body spectrum of Ref.~\cite{Horz:2019rrn} was recently analyzed in Ref.~\cite{Guo:2020kph} with a combination of variational approach and Faddeev formalism (see also Ref.~\cite{Guo:2018ibd}). Including relativistic kinematics and even effects of lattice spacing, the two and three-body spectra were qualitatively described in a one-parameter fit corresponding to the two-body interaction strength.

\subsection{Three-Body Force}

As noted above, a relevant output from  lattice calculations in the three-particle sector is the three-body force, which is defined differently in different approaches and is regularization-dependent within a given approach. The parameters of two-body interactions can be most conveniently extracted in the two-particle sector, and the interactions between more than three particles do not show up yet explicitly in the elastic energy window. The main problem in the extraction of the three-body force is that its contribution is very much suppressed, and a full control on the accuracy needs to be achieved, to separate this small effect from much larger contributions, coming from the two-body rescattering.

The first extraction of a three-body force from few-body systems in lattice QCD at higher than physical quark masses was performed by the NPLQCD collaboration in Ref.~\cite{Detmold:2008fn}. This paper also contained the pertinent lattice calculations for up to 12 pions at maximal isospin. In this work, the three-body force is parametrized by the coupling $\bar{\bar{\eta}}_3^L$ which was found to be non-zero except for the heaviest quark mass. Still, a direct comparison with non-perturbative approaches is difficult  -- prior to this, one needs to ``translate'' this result into the quantities  defined in these approaches. This can be done, performing the matching of the observable $S$-matrix elements, which must be the same in all approaches. In the context of this coupling constant, the relation to the threshold amplitude is discussed in detail in Ref.~\cite{Romero-Lopez:2020rdq}. 

The FVU formalism, discussed in Sec.~\ref{sec:3-body-fin-vol}
has also been applied~\cite{Mai:2018djl} to the analysis of the lattice data from Ref.~\cite{Detmold:2008fn}, extracting the three-body force $C_0$. Within the uncertainties and using a simple parametrization $C_0(\sqrt{s},\bm{p},\bm{q})=c_0\cdot\mathds{1}_{\bm{p}\bm{q}}$, this three-body force was found to be zero, $c_0= (0.3\pm 2.3)\times 10^{-6}$~MeV$^{-2}$. As mentioned above, this does not necessarily contradict the findings of Ref.~\cite{Detmold:2008fn}. Later, the excited levels of Ref.~\cite{Horz:2019rrn} for different boosts and irreps were analyzed with the RFT formalism, leading to a more precise determination of the three-body force~\cite{Blanton:2019vdk}. Technically, the two- and three-body
sectors were fitted jointly, with a simple parametrization for the energy dependence
of the two-body input adjusted to the corresponding two-body energy eigenvalues.
The three-body force was found to be non-zero. Even  attempts to determine its
energy dependence could be made, as the data of Ref.~\cite{Horz:2019rrn} are much
more precise and abundant than those of Ref.~\cite{Detmold:2008fn}. In addition,
the authors of Ref.~\cite{Blanton:2019vdk} calculated the three-to-three contact term
from LO ChPT and compared with the pertinent lattice values. While the
 energy-independent part was found to be non-zero and in broad agreement with the
LO ChPT prediction, the energy-dependent piece appeared in disagreement with that
prediction. 

Furthermore, the GWUQCD collaboration calculated the three-$\pi^+$ spectrum for
different quark masses, box geometries, and boosts, mapping out a plethora of states
and carrying out the comparison with the FVU predictions that were made
under the assumption of vanishing three-body forces~\cite{Culver:2019vvu}. A small subset of results is shown in Fig.~\ref{fig:schematic}. A fair agreement was found with a noticeable tension between the lattice data and predictions, leading to a $\chi^2_\text{dof}\approx 2.68$ or more,
depending on the two-body input. Yet, Ref.~\cite{Culver:2019vvu}
allowed, for the first time, to track the pion mass dependence of the three-body
amplitude and its qualitative agreement with the chiral extrapolation.

Subsequently, the ETMC collaboration calculated the three-$\pi^+$ spectrum at three
different pion masses, including the physical point for the first time~\cite{Fischer:2020jzp}. The extraction of the three-body force with the RFT formalism was compared with the LO ChPT prediction for the three-to-three process.
Similar to Ref.~\cite{Blanton:2019vdk}, the three-body term was found to be non-zero,
its energy-independent part being in qualitative agreement with LO ChPT, in contrast to the energy-dependent part.

Later, the three-body force was extracted by the
HadronSpectrum collaboration~\cite{Hansen:2020otl}, using the RFT formalism.
The pion mass in this calculation is relatively large ($M_\pi\approx 390$~MeV).
Within uncertainties and different fit strategies/parametrizations tried, the three-body
term in the isotropic approximation was found to be compatible with zero.  This study can be understood as a first step towards the production
of Dalitz plots from lattice QCD, because it included the actual solution of the
infinite-volume equivalent of the finite-volume three-body problem, with lattice input.
In Ref.~\cite{Hansen:2020otl}, some kinematical variables in the three-to-three amplitude were fixed, to be able to produce Dalitz plot-like distributions in the remaining variables. In this context, it should be mentioned that the FVU framework was recently extended
to the infinite volume (albeit without lattice input), in order to study the decay $a_1(1260)\to \pi\rho$ in coupled S- and D-waves.

Recently, the GWUQCD collaboration extracted the three-body force~\cite{Brett:2021wyd} from the data of Ref.~\cite{Culver:2019vvu}, using the FVU formalism~\cite{Brett:2021wyd}. Similarly to what was done in Ref.~\cite{Blanton:2019vdk}, the fit was performed without assuming any particular energy dependence for the three-body force. Each of the three-body energy eigenvalues was fitted individually, while leaving two-body input fixed. In general such a ``tomography plot'' allows one to map out energy and pion-mass dependence of the three-body force as shown in the right panel of Fig.~\ref{fig:chiralextrapolation-3b} in terms of a local three-body scattering amplitude $\bar T_3$, see Ref.~\cite{Brett:2021wyd} for more details and comparison to $C_0$ and $K_{\rm df,3}$. Then an energy-dependent global fit was also carried out, isolating the energy-independent and energy-dependent parts of the three-body force.

\begin{figure*}[t]
    \centering
    \includegraphics[width=0.495\textwidth,trim=0 0 0cm 0,clip]{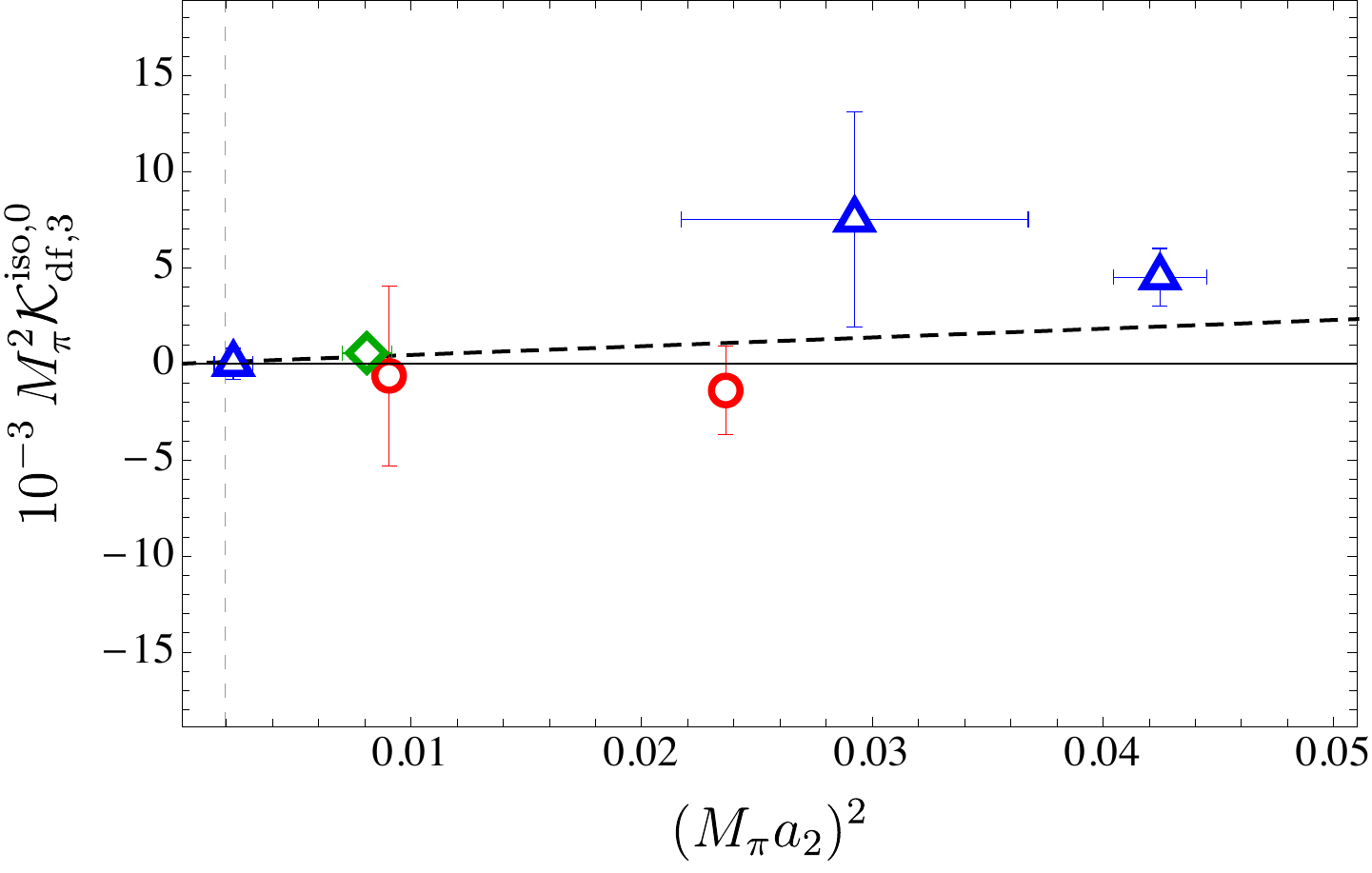}
    \includegraphics[width=0.495\textwidth,trim=0 0 0cm 0,clip]{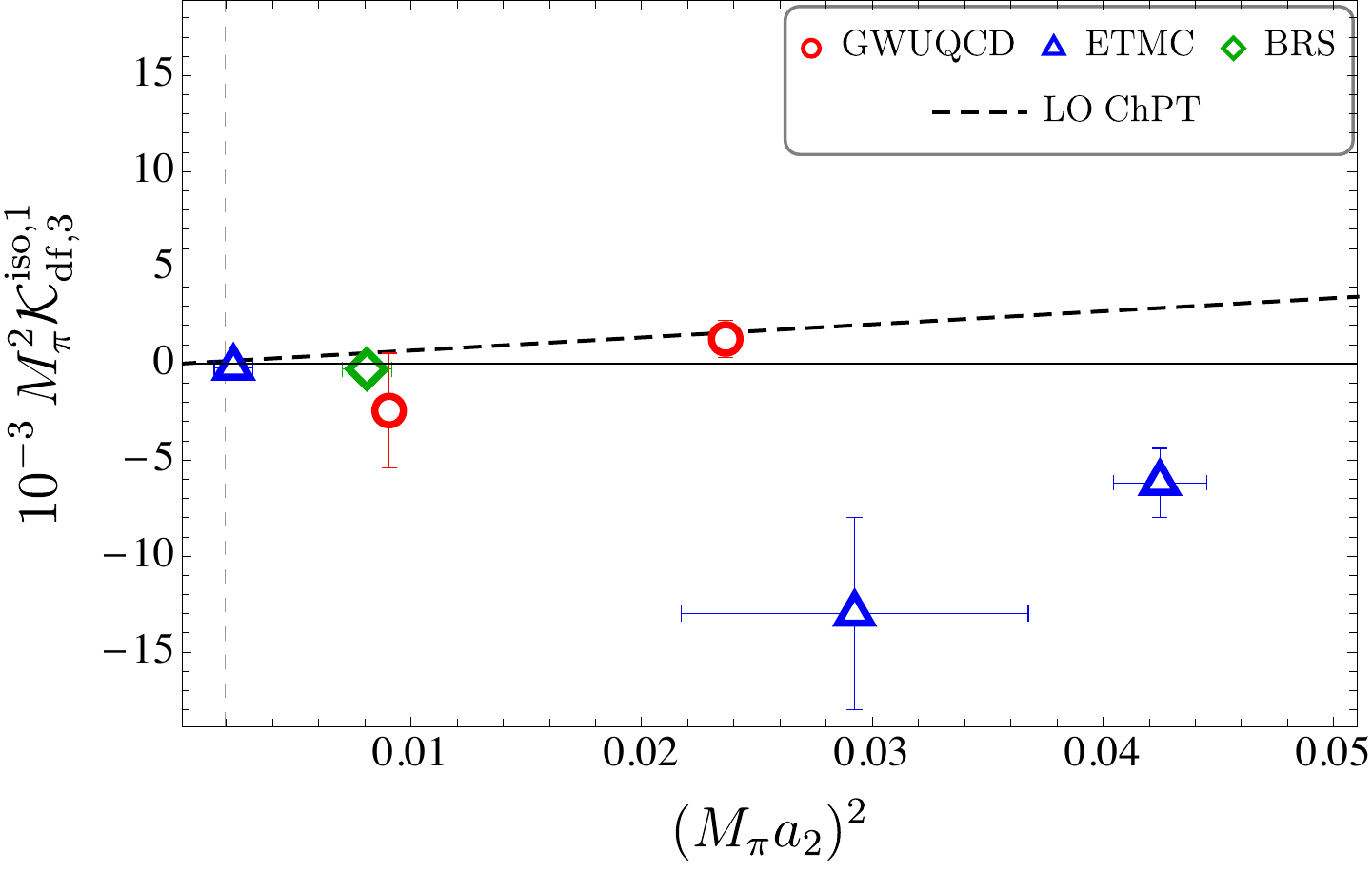}
    \caption{
      Three-body force $(I=3)$ as a function of the $I=2$ scattering length for the
      energy-independent (left) and energy-dependent (right) part. Results from
      GWUQCD/FVU~\cite{Brett:2021wyd}, ETMC~\cite{Fischer:2020jzp} and BRS~\cite{Blanton:2019vdk} are 
      shown by red circles, blue triangles and green diamonds, respectively. The leading order chiral prediction is denoted by the dashed line. The
      dashed vertical lines show the physical point.
    }
    \label{fig:T3barvsK3df}
\end{figure*}

A summary plot, taken from Ref.~\cite{Brett:2021wyd}, is provided in Fig.~\ref{fig:T3barvsK3df}. In particular, it shows the nominal value of the fitted three-body force in terms of the isotropic (spectator-momentum independent) parametrization $K_{\rm df,3}^{{\rm iso}}=K_{\rm df,3}^{{\rm iso},0}+(s/(9M_\pi^2)-1) K_{\rm df,3}^{{\rm iso}1}$ as a function of the $I=2$ scattering length. This representation allows one to compare dimensionless quantities, conveniently incorporating  the leading order chiral perturbation result, see Refs.~\cite{Blanton:2019vdk,Brett:2021wyd} for more details.
The overall picture is not entirely clear yet, in particular when comparing with the LO chiral predictions. The figure also shows that, at higher pion masses, there is, generally, a better chance to find a non-zero three-body force. Note, however, that there is no reason to expect that LO ChPT is valid for pion masses of up to three times the physical one.

Finally, we mention the extraction of the three-body force in $\varphi^4$ theory, which was carried out recently~\cite{Romero-Lopez:2018rcb,Romero-Lopez:2020rdq}. Even there is no direct
relation with the chiral extrapolation, these calculations, that were performed at many
different values of $L$, provide information which can be useful in the analysis of data
from lattice QCD. For example, it has been argued there that the effects of finite spacing might play an important role in the extraction of derivative (energy-dependent) couplings in the two-body sector, which are strongly correlated with the three-body force.

\subsection{Three kaons at maximal isospin}
\label{sec:threek}

To conclude this section, we briefly discuss extensions of three-body physics on
the lattice to the strangeness sector. As compared to the pion sector, there are only
few results focusing on the (multi-)strange
sector~\cite{Beane:2007uh,Sasaki:2013vxa,Helmes:2017smr}. The threshold energy
levels of multi-kaon states were first calculated by the NPLQCD
collaboration~\cite{Detmold:2008yn,Beane:2006kx} more than a decade ago. 
The first determination of excited levels was achieved by the GWUQCD collaboration in
Ref.~\cite{Alexandru:2020xqf} from ensembles that are generated with two
mass-degenerate light quarks ($N_f=2$ QCD), using the nHYP-smeared clover action.
The valence strange quark mass was tuned by setting the ratio $R=(M_K/M_{\pi})^2$
to its physical value.

Extending the FVU formalism of Refs.~\cite{Mai:2019fba, Mai:2018djl, Mai:2017bge} to the three-flavor sector allows for chiral extrapolations along arbitrary~$M_K(M_\pi)$ trajectories, using  constraints from chiral symmetry. Such implementations are standard in the two-body sector~\cite{Rendon:2020rtw, Molina:2020qpw, Hu:2017wli, Nebreda:2010wv, Pelaez:2010fj, Nebreda:2011di, Guo:2016zep}, but not yet explored for three-body systems.
The two-body input was extrapolated from the physical point, using the inverse
amplitude method and, in particular, the $SU(3)$ chiral amplitude to one
loop~\cite{GomezNicola:2001as,Truong:1988zp}. Low-energy constants were
chosen from a recent determination~\cite{Molina:2020qpw} that includes many
modern lattice QCD data along different $M_K(M_\pi)$ trajectories.

The NPLQCD lattice data for the $I=1$ $K K$ scattering length~\cite{Beane:2007uh}
and the three-body threshold energy shift~\cite{Beane:2007uh} were well predicted,
including their light quark-mass dependence  up to pion masses, for which chiral
extrapolations usually fail to converge ($M_\pi\approx 500$~MeV). Second, the excited
energy eigenvalues of the GWUQCD collaboration were, at least qualitatively, well
predicted, though there were some discrepancies at  higher energies.
These discrepancies might be a sign of a three-body force, but this question cannot
be settled without more lattice data. Also, it should be mentioned that the
elastic window for the $KKK$ system is particularly narrow due to the possibility of
 $\pi KKK$ states. This, again, points to the fact that unphysically heavy pions
can be advantageous in the study of multi-particle processes on the lattice.

\section{Decay into three-particle final states}
\label{sec:3-LL}

The latest developments of the three-particle formalism are not limited to the derivation of the quantization condition and the extraction of the three-particle force from lattice data. Recently, important progress has been achieved addressing decay processes with three-particle final states. Namely, a formula that relates the decay amplitudes in a finite and in the infinite volume, has been derived in the NREFT approach~\cite{Muller:2020wjo} and, later, in the RFT setting~\cite{Hansen:2021ofl}. In addition, a generalization to the case of non-rest frames, non-identical particles and  partial-wave mixing was discussed within the RFT approach. It is clear that the latest developments will boost the study of three-particle decays on the lattice. In this section, we give a brief overview of the recent developments in the field.

Generally, decay processes in QCD can be formally subdivided
into two categories. The decays of particles which are stable in pure QCD can be
attributed to the first category. This includes, for example, the weak decays $K\to 2\pi$, $K\to 3\pi$, but also electromagnetic transitions $\gamma^*\to 2\pi$, $\gamma^*\to 3\pi$ that
represent an important input  in the study of the muon $g-2$ factor.
Moreover, the decays that occur only when isospin is not conserved, as e.g.,
in $\eta\to 3\pi$, can be also included here (in this case, a particle is stable in
pure QCD with equal quark masses, $m_d=m_u$, and the decay amplitude is
proportional to $(m_d-m_u)$. Since the interactions that lead to such decays are much
weaker than strong interactions, they can be considered at the first order in
perturbation theory. The particles in the unperturbed theory are stable and their
masses are located on the real energy axis. In contrast to this, the strong decays,
like $\rho\to 2\pi$ or $\Delta\to N\pi$, belong to the second category. Such unstable
particles correspond to poles in the complex energy plane. In order to extract the
parameters of these decays on the lattice (the real and imaginary parts of the pole
position), as well as the matrix elements containing these resonances (say, the
electromagnetic form factor of a resonance), one has to perform analytic continuation
of lattice data from the real axis to a resonance pole.

Similar to the two- and three-particle scattering amplitudes, discussed above,
the decay amplitudes obtained from a finite-volume calculation cannot  be
simply mapped onto the physical ones in  the limit $L\to\infty$. Again, this non-trivial
volume dependence can be attributed to the three-body final-state interaction.
Hence, in order to be able to interpret the lattice results, one has to first derive a
formula, which relates the amplitudes in a finite and in the infinite volume. A crucial
point is that the final-state interactions represent a long-range phenomenon and,
therefore, one may utilize the effective field theories of QCD in order to arrive at
the desired result.

In their seminal paper, Lellouch and L\"uscher~\cite{Lellouch:2000pv} have shown that the finite- and infinite-volume matrix elements for the $K\to 2\pi$ decay are related
by a single factor (LL factor), which depends only on the $\pi\pi$ 
phase shift and $L$. The LL factor contains all power-law $L$-dependence. Thus,
removing this factor, one may perform the limit $L\to\infty$. The absolute value of
the infinite-volume matrix element is obtained in this limit, and the phase of this matrix
element, which is determined by Watson's theorem, can be also measured on the same
lattice configuration. Note also that $K\to 2\pi$ belongs to the first category, and hence
the analytic continuation should not be considered.

The paper~\cite{Lellouch:2000pv}
paved the way to the systematic investigation of two-body decays on the lattice.
A comprehensive study of the $K\to\pi\pi$ decays, carried out recently by the
RBC and UKQCD Collaborations~\cite{Abbott:2020hxn}, is just one example of this. 
Various generalizations of the Lellouch-L\"uscher approach emerged.
In particular, it has been extended to  moving frames~\cite{Kim:2005gf,Christ:2005gi}
and  coupled two-body channels~\cite{Hansen:2012tf}. A simple and transparent
derivation of the Lellouch-L\"uscher formula with the use of NREFT
formalism has been given in~\cite{Bernard:2012bi} where, in particular, the analytic
continuation to the resonance pole is discussed in detail, see also
Refs.~\cite{Agadjanov:2014kha,Agadjanov:2016fbd}.
In relation to this work we also mention the study of matrix elements of currents,
corresponding to the $1\to 2$ transition~\cite{Briceno:2015csa,Briceno:2014uqa}, and 
of the timelike pion form factor~\cite{Meyer:2011um}, which all feature the LL factor in a
finite volume.

Despite significant progress in the description of two-body decays, the
decays into three particles have remained {\it terra incognita} until very
recently. The crucial difference between two- and three-particle cases consists
in the fact that, in the two-particle system, there exists a single LL factor
that relates the matrix elements in a finite and in the infinite volume. This
is easy to understand from kinematics alone. Indeed, in the center-of-mass
frame, the magnitude of the momenta of the decay products is determined by the
mass of the decaying particle. Hence, there is no variable left for the LL
factor to depend on. On the contrary, in three-particle decays, the relative
momenta are not completely fixed by energy-momentum conservation and
the LL factor in general depends on the momenta. For this reason,
in order to extract the matrix element, it is convenient to adopt a two-step approach. First, the momentum dependence of the short-range part of the decay vertex should
be parametrized, e.g., via  polynomials of a given order. This parametrization is
already built in the NREFT approach~\cite{Muller:2020wjo}, and can be conveniently
introduced in the RFT setting~\cite{Hansen:2021ofl}, expanding the vertex in the vicinity
of the decay threshold. Second, the long-range part can be systematically calculated
(in both approaches) within effective theory in a finite volume, leading to the
momentum-dependent LL factor one is looking for. This LL factor depends on the
parameters of the interactions in the final state (to be measured simultaneously with the matrix element), but not on the interactions that lead to a particular decay. The information about the latter is contained solely in the coefficients
of the above-mentioned polynomial that should be fitted to the results of the
lattice measurements at different momenta of external particles. Finally,
the physical matrix element is obtained by combining this polynomial
with the long-range part, which is determined by the solutions of an
infinite-volume integral equation that describes scattering in a final state.
The procedure described above represents an analog of the LL formalism
for a three-particle system.

\section{Conclusions and outlook}
\label{sec:concl}

Recent years have seen a surge of interest in the study of three- (and more) particle  systems on the lattice. On the side of formalism, a major breakthrough
was associated with the derivation of the quantization condition, which relates the finite-volume spectrum with $S$-matrix elements in the three-particle sector. Several physically equivalent formulations of the quantization condition are available at present.
In addition, a three-particle analog of the Lellouch-L\"uscher formula has been derived
recently, which enables one to perform lattice measurements of three-particle decay
matrix elements. These developments have boosted lattice simulations in the three-particle
sector which, at present, are mainly focused on the extraction of the three-particle
force from the finite-volume energy levels.

The advance of the lattice studies may have far-reaching implications in particle and
nuclear physics. First and foremost, this concerns three-particle processes in the light
quark sector, say, the decays of charged and neutral kaons,
$\eta,\eta',\omega$ and $a_1(1260)$ mesons~\cite{Sadasivan:2020syi}. Here, one should also mention the
long-standing problem with the Roper resonance~\cite{Lang:2016hnn}, which decays with a significant fraction
into the $\pi\pi N$ channel. There are very interesting applications
in the charm-quark sector as well, for example, in the study of the process
$X(3872)\to D\bar D^*\to D\bar D\pi$~\cite{Padmanath:2015era, Baru:2013rta,  Garzon:2013uwa}. 
As for other multi-meson systems, note that the study of pion, kaon, and proton correlations in heavy ion collisions
by the ALICE@CERN collaboration~\cite{Adam:2015vja} relies on the value of the
$K^-K^-$ scattering length determined in a lattice calculation~\cite{Beane:2007uh}.
Apart from this, the information about multi-$K^-$ systems is relevant for the
understanding of strange nuclear matter and its implications to the equation of state
of neutron stars. In particular, it is well known that ultra-dense environments (such as
those in the core of neutron stars) allow for an appearance of kaon
condensates~\cite{Kaplan:1986yq,Li:1997zb,Pal:2000pb,Lee:1996ef}, that can soften
the equation of state of neutron
stars~\cite{Li:1997zb,Pal:2000pb,Lonardoni:2014bwa,Hell:2014xva}
(further details on the antikaon interaction with baryonic matter can be found in
reviews~\cite{Gal:2016boi,Mai:2020ltx}). In this context, quantifying multineutron forces is also necessary for the
equation of state of neutron matter in the extreme conditions
of a neutron star~\cite{Savage:2016egr}. Recent advances in lattice QCD
on few-nucleon systems~\cite{Horz:2020zvv, Orginos:2015aya} complement dedicated experimental
programs, e.g., at the FRIB facility~\cite{Gade:2016xrp}.

In view of such perspectives, a question about the optimal strategies
for carrying out lattice calculations in multi-particle systems becomes important. In the present review, we have focused in particular on the issue of chiral extrapolations.
There are several reasons to do this. First of all, the lattice calculations at present are often
carried out at larger than physical quark masses. This will remain so in a foreseeable future, especially in the multi-particle sector. Hence, one has to learn to perform a global
fit to {\em all} available data, taken at different quark masses, and even combine them with  experimental data, in order to reliably
extract the physical quantities of interest. This cannot be
achieved without a robust control over the chiral extrapolations that is provided by the use of the (unitarized versions of) ChPT. 
Furthermore, for quark masses close to physical ones, inelastic channels come  close and there are only few data points available in the elastic region, if the volume is not taken very large. Finally, it should be also noted that, for large quark masses, some of the light resonances in the $\pi\pi$ scattering (e.g., $f_0(500),\,\rho(770)$) become bound states, see, e.g.,~\cite{Hanhart:2008mx, Briceno:2016mjc, Doring:2016bdr, Guo:2018zss}. Hence, for larger quark masses and energies below the breakup threshold, the three-particle problem effectively turns into a two-particle one and, in some cases, it might be possible
to apply the L\"uscher and Lellouch-L\"uscher formalisms, combined with chiral extrapolation, in order to extract three-particle observables.

\begin{acknowledgement}
\begin{sloppypar}
We thank R.~Brett for a careful reading of the manuscript.
The work of M.D. and M.M. is supported by the National  Science  Foundation  under  Grant  No.   PHY-2012289 and by the U.S. Department of Energy under Award No. DE-SC0016582. M.D. is also supported by the U.S. Department of Energy, Office of Science, Office of Nuclear Physics under contract DE-AC05-06OR23177. The work of A.R.  is  funded in part by the Deutsche Forschungsgemeinschaft
(DFG, German Research Foundation) – Project-ID 196253076 – TRR 110,
Volkswagenstiftung  (Grant No. 93562) and the Chinese Academy of Sciences
(CAS) President's International Fellowship Initiative (PIFI) (Grant No. 2021VMB0007).
\end{sloppypar}
\end{acknowledgement}

\bibliography{BIB}

\end{document}